\documentstyle[12pt]{article}

\newcounter{eq}
\newcounter{sc}

\newcommand {\PL}   {Phys. Lett.}

\def\overleftrightarrow#1{\vbox{\ialign{##\crcr
 $\leftrightarrow$\crcr\noalign{\kern-1pt\nointerlineskip}
 $\hfil\displaystyle{#1}\hfil$\crcr}}}

\newlength{\minitwocolumn}
\begin{document}

\begin{flushright}
DPUR/TH/25\\
July, 2011\\
\end{flushright}
\vspace{20pt}

\pagestyle{empty}
\baselineskip15pt

\begin{center}
{\large\bf Emergence of Superstring from Pure Spinor
\vskip 1mm }

\vspace{20mm}
Ichiro Oda
          \footnote{
          E-mail address:\ ioda@phys.u-ryukyu.ac.jp} \\

\vspace{5mm}
           Department of Physics, Faculty of Science, University of the 
           Ryukyus,\\
           Nishihara, Okinawa 903-0213, Japan.\\

\end{center}


\vspace{5mm}
\begin{abstract}
Starting with a classical action where a pure spinor $\lambda^\alpha$ is only a
fundamental and dynamical variable, the pure spinor formalism for 
superstring is derived by following the BRST formalism.
In this formalism, not only the string variable $x^m$ but also the space-time spinor 
$\theta^\alpha$ are emerged as the Faddeev-Popov (FP) ghosts of a topological symmetry 
and its reducible symmetry. This study suggests that the fundamental theory behind the
pure spinor formalism of the superstring might be a topological field theory. 
\end{abstract}

\newpage
\pagestyle{plain}
\pagenumbering{arabic}


\section{\large\bf Introduction}

It is interesting to inquire where superstring theory, which has been
considered as a promising candidate of theory of everything, comes from
and dream that it might be emerged from a quite trivial theory like a
topological theory as in the scenario of creation of universes from nothing in
cosmology. One of motivations in this article is to pursue such an
idea and to suggest that the superstring in the pure spinor formalism 
\cite{Ber1} - \cite{Luca} might be constructed out of a class of topological field 
theories \cite{Witten}.

Relevantly to our motivation, Berkovits has recently advocated a new interpretation 
of the BRST charge $Q_B$ in the pure spinor formalism of the superstring \cite{Ber0}.
In this new interpretation, instead of regarding $(x^m, \theta^\alpha)$ 
and $\lambda^\alpha$ as matter and ghost variables, respectively, 
$(x^m, \lambda^\alpha)$ and $\theta^\alpha$ are regarded as matter and 
ghost variables. The matter variables then satisfy a twistor-like constraint 
instead of the Virasoro constraint. It turns out that quantizing
this twistor-like constraint yields the fermionic Faddeev-Popov (FP) ghost $\theta^\alpha$ 
and the nilpotent BRST charge. After twisting the ghost number, it is shown that
the BRST cohomology is related to the cohomology of the pure spinor formalism
of the superstring. It is of interest to note that the fermionic coordinate
$\theta^\alpha$ is emerged as the FP ghost via the standard BRST quantization 
procedure.

It is then natural to ask ourselves if in addition to the fermionic coordinate
$\theta^\alpha$ the bosonic string coordinate $x^m$ could be emerged in a
similar manner since the bosonic string coordinate $x^m$ is on the same footing
as the fermionic coordinate $\theta^\alpha$ in a supersymmetric theory.

In this article, we would like to propose such a formalism where 
only the pure spinor $\lambda^\alpha$ is a dynamical variable while supersymmetric
string coordinates $(x^m, \theta^\alpha)$ are emerged as the FP ghosts via the
BRST formalism. In our approach, it is remarkable that starting with a trivial 
action of topological quantum field theory, the superstring coordinates $(x^m, \theta^\alpha)$
are appeared only at the quantum level through the gauge-fixing of a topological symmetry
and its reducible symmetry. 
In this sense, the origin of the pure spinor formalism of the superstring might be
a topological field theory. 

The idea that the pure spinor formalism of the superstring stems from a topological field theory 
is supported by counting degrees of freedom in the both theories. Namely,
it is well-known that topological field theories possess an equal number of bosonic
and fermionic degrees of freedom. Thus, if the pure spinor formalism of the superstring is somehow
derived from a topological field theory, the both theories should have the same number of bosonic
and fermionic degrees of freedom. Indeed, the pure spinor formalism of the superstring has
$32$ bosonic and $32$ fermionic degrees of freedom, therby giving us a $c=0$ conformal field
theory as in a topological field theory. Accordingly, there could be a possibility that 
the pure spinor formalism of the superstring has an origin of a topological field theory.

\section{\large\bf Superparticle}

Before discussing a case of the superstring, it is worth investigating 
a case of the superparticle in ten dimensions even if the BRST-invariant
action is slightly distinct from the usual superparticle action in the
pure spinor formalism in that only $5$ independent components of $x^m$
appear in the action.

We start with the following superparticle action in ten dimensions:
\begin{eqnarray}
S_c = \int d \tau ( \omega_\alpha \dot{\lambda}^\alpha 
+ f_\alpha \lambda^\alpha ),
\label{SP action1}
\end{eqnarray}
where the spinor index $\alpha$ runs from $1$ to $16$ (and the vector index $m$, 
which appears below, runs from $0$ to $9$), and the dot denotes a derivative 
with respect to $\tau$. $\lambda^\alpha$ is a bosonic pure spinor variable of 
ghost number $1$ satisfying the pure spinor condition
\begin{eqnarray}
\lambda^\alpha \gamma^m_{\alpha\beta} \lambda^\beta = 0.
\label{PS condition}
\end{eqnarray}

Now let us perform the canonical analysis of the action (\ref{SP action1}). 
The canonical conjugate momenta are defined as
\begin{eqnarray}
\omega_\alpha &=& \frac{\partial S_c}{\partial \dot{\lambda}^\alpha}, \nonumber\\
\pi^\alpha &=& \frac{\partial S_c}{\partial \dot{f}_\alpha} \approx 0.
\label{Momenta}
\end{eqnarray}
The last equality is a primary constraint. The Hamiltonian $H$ is then
of form
\begin{eqnarray}
H = - f_\alpha \lambda^\alpha.
\label{Hamiltonian}
\end{eqnarray}
Using this Hamiltonian, the time development of the primary
constraint leads to a secondary constraint
\begin{eqnarray}
\lambda^\alpha \approx 0.
\label{2nd-constraint}
\end{eqnarray}
It is easy to see that there is no ternary constraint and these constraints
constitute the first-class constraints. Incidentally, the secondary constraint
(\ref{2nd-constraint}) renders the classical action (\ref{SP action1}) vanishing,
thereby implying a topological nature of the classical theory at hand.
Namely, the action (\ref{SP action1}) belongs to the Witten type of topological 
field theories \cite{Witten}.

The generator of a topological symmetry takes the form \cite{Oda3}
\begin{eqnarray}
G = - \dot{\varepsilon}_\alpha \pi^\alpha + \varepsilon_\alpha \lambda^\alpha,
\label{Generator}
\end{eqnarray}
where $\varepsilon_\alpha$ is a bosonic local parameter. 
With this generator, the topological symmetry reads
\begin{eqnarray}
\delta \omega_\alpha &=& \varepsilon_\alpha, \nonumber\\
\delta f_\alpha &=& \dot{\varepsilon}_\alpha, \nonumber\\
\delta \lambda^\alpha &=& 0. 
\label{Topological symmetry}
\end{eqnarray}
Actually, the classical action (\ref{SP action1}) is invariant under this symmetry
up to a surface term
\begin{eqnarray}
\delta S_c = \int d \tau \frac{d}{d \tau} ( \varepsilon_\alpha \lambda^\alpha).
\label{delta S}
\end{eqnarray}

By replacing the bosonic parameter $\varepsilon_\alpha$ with 
the fermionic ghost $p_\alpha$ of ghost number $0$, one obtains the BRST transformation
associated with the topological symmetry as follows:    
\begin{eqnarray}
\delta_B \omega_\alpha &=& p_\alpha, \nonumber\\
\delta_B f_\alpha &=& \dot{p}_\alpha, \nonumber\\
\delta_B \theta^\alpha &=& - b^\alpha, \nonumber\\
\delta_B p_\alpha &=& \delta_B b^\alpha = 0, 
\label{BRST1}
\end{eqnarray}
where $\theta^\alpha$ is a fermionic antighost of ghost number $0$
and $b^\alpha$ is a bosonic auxiliary field of ghost number $1$.

We shall fix the topological symmetry by a gauge condition
$f_\alpha = 0$, so that the gauge fermion is $\Psi_1 = - \theta^\alpha f_\alpha$. 
However, only 11 of 16 components of the secondary constraint (\ref{2nd-constraint})
are independent, so we still have a reducible symmetry $\delta f_\alpha 
= \epsilon_\alpha$ of 5 components satisfying $\epsilon_\alpha \lambda^\alpha
= \epsilon \gamma^{mn} \lambda = 0$.
To treat this reducible symmetry in an appropriate manner, we introduce bosonic ghosts
for ghosts $u_\alpha$, which have ghost number $1$ and 5 independent components, such that
$u_\alpha \lambda^\alpha = u \gamma^{mn} \lambda = 0$. 
Here the BRST transformation reads   
\begin{eqnarray}
\delta_B p_\alpha &=& u_\alpha, \nonumber\\
\delta_B v^\alpha &=& B^\alpha, \nonumber\\
\delta_B u_\alpha &=& \delta_B B^\alpha = 0, 
\label{Reducible1}
\end{eqnarray}
where $v^\alpha$ is a bosonic antighost of ghost number $- 1$ and $B^\alpha$ 
is a fermionic auxiliary field of ghost number $0$.
To fix this reducible symmetry, we take a gauge condition $\dot{p}_\alpha = 0$
so that the gauge fermion becomes $\Psi_2 = v^\alpha \dot{p}_\alpha$. 

Consequently, we have a gauge-fixed, BRST-invariant action
\begin{eqnarray}
S &\equiv& S_c + \int d \tau \delta_B (\Psi_1 + \Psi_2)   \nonumber\\
&=& \int d \tau ( \omega_\alpha \dot{\lambda}^\alpha 
+ p_\alpha \dot{B}^{\prime \alpha} + f_\alpha b^{\prime \alpha} 
+ v^\alpha \dot{u}_\alpha ),
\label{SP1}
\end{eqnarray}
where we have defined $B^{\prime \alpha} = B^\alpha + \theta^\alpha$ and
$b^{\prime \alpha} = b^\alpha + \lambda^\alpha$. Then, after integrating over 
$f_\alpha, b^{\prime \alpha}$ and rewriting $B^{\prime \alpha}$ as $\theta^\alpha$, 
we arrive at a quantum action
\begin{eqnarray}
S = \int d \tau ( \omega_\alpha \dot{\lambda}^\alpha 
+ p_\alpha \dot{\theta}^\alpha + v^\alpha \dot{u}_\alpha ).
\label{SP2}
\end{eqnarray}
Indeed, this action is invariant under the BRST transformation 
up to a surface term
\begin{eqnarray}
\delta_B S = \int d \tau \frac{d}{d \tau} ( u_\alpha \theta^\alpha).
\label{BRST2}
\end{eqnarray}
Let us note that at this stage the BRST transformation is reduced to 
the form
\begin{eqnarray}
\delta_B \omega_\alpha &=& p_\alpha, \nonumber\\
\delta_B \theta^\alpha &=& \lambda^\alpha, \nonumber\\
\delta_B p_\alpha &=& u_\alpha, \nonumber\\
\delta_B v^\alpha &=& \theta^\alpha, \nonumber\\
\delta_B \lambda^\alpha &=& \delta_B u_\alpha = 0. 
\label{BRST2-2}
\end{eqnarray}
The BRST charge then takes the form 
\begin{eqnarray}
Q_B = \lambda^\alpha p_\alpha + u_\alpha \theta^\alpha.
\label{BRST charge1}
\end{eqnarray}

To verify that this BRST charge is related to that of the superparticle in the 
pure spinor formalism, it is enough to notice that $u_\alpha$ can be described in
terms of a space-time vector $P_m$ as 
\begin{eqnarray}
u_\alpha = P_m (\gamma^m \lambda)_\alpha,
\label{u1}
\end{eqnarray}
since the both sides have $5$ independent components. 
Then the BRST charge (\ref{BRST charge1}) can be rewritten as  
\begin{eqnarray}
Q_B &=& \lambda^\alpha p_\alpha + P_m (\lambda \gamma^m  \theta) \nonumber\\
&=& \lambda^\alpha [ p_\alpha + P_m (\gamma^m  \theta)_\alpha ] \nonumber\\
&=& \lambda^\alpha [ \partial_\alpha - i (\gamma^m  \theta)_\alpha \partial_m ] 
\nonumber\\
&\equiv& \lambda^\alpha D_\alpha,
\label{BRST charge2}
\end{eqnarray}
where $D_\alpha$ is the supersymmetric derivative, and we have set $p_\alpha 
= \frac{\partial}{\partial \theta^\alpha} \equiv \partial_\alpha$
and $P_m = - i \frac{\partial}{\partial x^m} \equiv - i \partial_m$. 
 
Furthermore, with the definition of $x^m \equiv - (v \gamma^m \lambda)$, 
which has ghost number $0$, the action (\ref{SP2}) is cast to the form
\begin{eqnarray}
S = \int d \tau ( \omega^\prime_\alpha \dot{\lambda}^\alpha 
+ p_\alpha \dot{\theta}^\alpha + P_m \dot{x}^m ),
\label{SP3}
\end{eqnarray}
where we have defined $\omega^\prime_\alpha \equiv \omega_\alpha 
+ P_m ( \gamma^m v)_\alpha$. If we rewrite $\omega^\prime_\alpha$
as  $\omega_\alpha$, we finally have a BRST-invariant action for the
superparticle
\begin{eqnarray}
S = \int d \tau ( \omega_\alpha \dot{\lambda}^\alpha 
+ p_\alpha \dot{\theta}^\alpha + P_m \dot{x}^m ).
\label{SP4}
\end{eqnarray}
Then, the BRST transformation is given by
\begin{eqnarray}
\delta_B \omega_\alpha &=& p_\alpha + P_m (\gamma^m \theta)_\alpha, \nonumber\\
\delta_B p_\alpha &=& P_m (\gamma^m \lambda)_\alpha, \nonumber\\
\delta_B x^m &=& - ( \theta \gamma^m \lambda), \nonumber\\
\delta_B \theta^\alpha &=& \lambda^\alpha, \nonumber\\
\delta_B P_m &=& \delta_B \lambda^\alpha = 0.
\label{BRST3}
\end{eqnarray}

At first sight, it might appear that we have exactly obtained the BRST-invariant action 
for the superparticle in the pure spinor formalism, but this is an illusion,
whose reason we will explain below in two different ways.

First let us note that our definition of $x^m \equiv - (v \gamma^m \lambda)$ yields 
a null constraint $x^m x_m = 0$. 
The root of this problem is traced to the fact that our $x^m$ satisfies an equation
\begin{eqnarray}
x^m ( \gamma_m \lambda )_\alpha = 0.
\label{xm}
\end{eqnarray}
Here recall that in the $U(5)$ decomposition of $SO(10)$, a space-time vector $y^m$ and 
space-time spinor $f^\alpha$ are described as $y^m = y^a \oplus y_a \in 5 \oplus \bar 5$ 
and $f^\alpha = f^{+} \oplus f_{[ab]} \oplus f^a \in 1 \oplus \bar {10} \oplus 5$ 
where the indices $a, b$ take the values $1, \cdots, 5$. 
Using the $U(5)$ decomposition,  Eq. (\ref{xm}) is divided into three equations
\begin{eqnarray}
x_b \lambda^{ba} + x^a \lambda^+ &=& 0, \nonumber\\
( - \frac{1}{3!} \varepsilon^{abcde} \lambda_b x_a 
+ \frac{1}{2} \lambda^{cd} x^e ) \varepsilon_{cdefg} &=& 0, \nonumber\\
x^a \lambda_a &=& 0.
\label{xm2}
\end{eqnarray}
The general solution for (\ref{xm2}) turns out to be
\begin{eqnarray}
x^a = \frac{1}{\lambda^+} \lambda^{ab} x_b,
\label{xm3}
\end{eqnarray}
so our $x^m$ has only $5$ independent components, which is $x_a \in \bar 5$.
Thus, the last term in the action (\ref{SP4}) should be written as 
$P^a \dot x_a$ instead of $P_m \dot{x}^m$.

Next, let us show the same fact by counting the independent degrees of freedom
of variables. We have started with a topological theory (\ref{SP action1}).
It is known that topological field theories have an equal number of bosonic and fermionic 
degrees of freedom, so the action (\ref{SP action1}) should share such a feature. 
In fact, in the action (\ref{SP4}) where $P_m \dot{x}^m$ is replaced
with $P^a \dot x_a$, as bosonic degrees of freedom, we have $11 \omega_\alpha,
11 \lambda^\alpha, 5 P^a, 5 x_a$ so that we have in total 32 while as fermionic
degrees of freedom, we have $16 p_\alpha, 16 \theta^\alpha$ so that we have totally 32.
Thus it is certain that as required by topological field theories we have the same number 
of bosonic and fermionic degrees of freedom in the action (\ref{SP4}) if we replace 
$P_m \dot{x}^m$ with $P^a \dot x_a$.

To close this section, it is valuable to point out that all the variables
have proper ghost number assignment without twisting the ghost number.
Of course, our ghost number assignment can be read out from a scale invariance
of the action (\ref{SP4})
\begin{eqnarray}
P_m &\rightarrow& P_m, \ x^m \rightarrow x^m, \
\omega_\alpha \rightarrow e^{-\rho} \omega_\alpha, \ 
\lambda^\alpha \rightarrow e^\rho \lambda^\alpha, 
\nonumber\\
p_\alpha &\rightarrow& p_\alpha, \ \theta^\alpha \rightarrow \theta^\alpha.
\label{Scale}
\end{eqnarray}
With this scale transformation, the ghost number can be defined to
each variable as 
\begin{eqnarray}
x^m (0), P_m (0), \theta^\alpha (0), p_\alpha (0),
\lambda^\alpha (1), \omega_\alpha (-1),
\label{Ghost number}
\end{eqnarray}
where the values in the curly bracket after variables denote the ghost
number. Note that as a result the BRST charge $Q_B$ in (\ref{BRST charge2}) has 
ghost number $1$ as desired.

\section{\large\bf Superstring}

In this section, we move on to the superstring in ten dimensions, which is the
main part of this article. A classical action for the superstring 
on the world-sheet is made out of the left and right-moving bosonic variables 
$\lambda^\alpha$ and $\hat{\lambda}^{\hat{\alpha}}$ satisfying the pure spinor conditions 
$\lambda \gamma^m \lambda = \hat{\lambda} \gamma^m \hat{\lambda} = 0$ as follows:
\begin{eqnarray}
S_c = \int d^2 z ( \omega_\alpha \bar \partial \lambda^\alpha 
+ \hat{\omega}_{\hat{\alpha}} \partial \hat{\lambda}^{\hat{\alpha}} 
+ f_\alpha \lambda^\alpha + \hat{f}_{\hat{\alpha}} \hat{\lambda}^{\hat{\alpha}} ).
\label{ST action1}
\end{eqnarray}
Here $\lambda^\alpha$ and $\hat{\lambda}^{\hat{\alpha}}$ have the same space-time chirality 
for the Type IIB superstring and the opposite space-time chirality for the Type IIA superstring.
For simplicity, we shall confine ourselves to only the left-moving (holomorphic)
sector of a closed superstring since the generalization to the right-moving
(anti-holomorphic) sector is straightforward.

According to a perfectly similar line of the argument to the superparticle, it is easy to
obtain the following BRST-invariant action for the superstring
\begin{eqnarray}
S = \int d^2 z ( \omega_\alpha \bar \partial \lambda^\alpha 
+ p_\alpha \bar \partial \theta^\alpha + P^a \bar \partial x_a ).
\label{ST1}
\end{eqnarray}
In order to have the superstring action in the pure spinor formalism,
it is sufficient to choose 
\begin{eqnarray}
P^a = \partial x^a,
\label{Pa}
\end{eqnarray}
and then substitute it into the action (\ref{ST1}) whose result reads
\begin{eqnarray}
S = \int d^2 z ( \frac{1}{2} \partial x^m \bar \partial x_m 
+ p_\alpha \bar \partial \theta^\alpha 
+ \omega_\alpha \bar \partial \lambda^\alpha ),
\label{ST2}
\end{eqnarray}
where we have used the relation 
\begin{eqnarray}
\partial x_a \bar \partial x^a + \partial x^a \bar \partial x_a
= \partial x^m \bar \partial x_m.
\label{Px}
\end{eqnarray}

This action is BRST-invariant under the BRST transformation generated
by the BRST charge $Q_B$ for the superstring
\begin{eqnarray}
Q_B = \oint d z \lambda^\alpha d_\alpha, 
\label{S-BRST charge}
\end{eqnarray}
where $d_\alpha$ is the supersymmetric variable defined as \footnote{The term
$- \frac{1}{2} (\theta \gamma^m \partial \theta) (\gamma_m \theta)_\alpha$ is
added to $d_\alpha$ to compensate for the choice (\ref{Pa}) in such a way that
the BRST charge $Q_B$ becomes nilpotent using the pure spinor condition.}
\begin{eqnarray}
d_\alpha 
\equiv p_\alpha + [ \partial x^m - \frac{1}{2} (\theta \gamma^m \partial \theta) ]
(\gamma_m \theta)_\alpha.
\label{d}
\end{eqnarray}
For instance, the BRST transformation is given by
\begin{eqnarray}
\delta_B \omega_\alpha &=& d_\alpha, \nonumber\\
\delta_B x^m &=& - (\lambda \gamma^m \theta), \nonumber\\
\delta_B \theta^\alpha &=& \lambda^\alpha.
\label{S-BRST}
\end{eqnarray}

Here it is valuable to mention that in contrast to the superparticle, 
we have precisely obtained the superstring in the pure spinor formalism. 
This fact is checked by counting the independent degrees of freedom as follows: 
Since we have started with a topological action (\ref{ST action1}), the BRST-invariant 
action (\ref{ST2}) should have an equal number of bosonic and fermionic 
degrees of freedom. Actually, as bosonic degrees of freedom, we have $11 \omega_\alpha,
11 \lambda^\alpha, 10 x^m$ so that we have in total 32 while as fermionic
degrees of freedom, we have $16 p_\alpha, 16 \theta^\alpha$ so that we have totally 32.

Finally, let us mention that the physical state condition for the Type II closed superstring 
is provided by $Q_B \Phi = \hat{Q}_B \Phi = 0$ for physical states $\Phi$ where 
$\Phi$ is a functional of $x^m, \theta^\alpha, \hat{\theta}^{\hat{\alpha}},
\lambda^\alpha, \hat{\lambda}^{\hat{\alpha}}$, that is, $\Phi 
= \Phi( x^m, \theta^\alpha, \hat{\theta}^{\hat{\alpha}}, \lambda^\alpha, \hat{\lambda}^{\hat{\alpha}} )$. 
Given that a massless vertex operator 
$\Phi = \lambda^\alpha \hat{\lambda}^{\hat{\beta}} A_{\alpha \hat{\beta}}(x, \theta, \hat{\theta})$ 
with a superfield $A_{\alpha \hat{\beta}}(x, \theta, \hat{\theta})$, the physical state condition yields 
the Type II supergravity equations 
$ \gamma_{m_1 \cdots m_5}^{\alpha\beta} D_\alpha A_{\beta \hat{\gamma}} 
= \gamma_{m_1 \cdots m_5}^{\hat{\alpha} \hat{\gamma}} \hat{D}_{\hat{\alpha}} A_{\beta \hat{\gamma}} = 0$,
and $\delta \Phi = Q_B \hat{\Omega} + \hat{Q}_B \Omega$ gives us the gauge transformation 
of the Type II supergravity, 
$\delta A_{\alpha \hat{\beta}} = D_\alpha \hat{\Omega}_{\hat{\beta}} 
+ \hat{D}_{\hat{\beta}} \Omega_\alpha,
\gamma_{m_1 \cdots m_5}^{\alpha\beta} D_\alpha \Omega_\beta 
= \gamma_{m_1 \cdots m_5}^{\hat{\alpha} \hat{\gamma}} \hat{D}_{\hat{\alpha}} \hat{\Omega}_{\hat{\gamma}} = 0$. 
Thus, our BRST cohomology for massless sector implies the Type II supergravity theory. 
In a similar manner, we can define the massive physical states.

\section{\large\bf Conclusion}

In this article, on the basis of the BRST quantization procedure,
we have derived the superstring in the pure spinor formalism by starting 
with a classical action composed of only the pure spinor $\lambda^\alpha$ 
as a dynamical variable. In our
approach, the supersymmetric coordinates $( x^m, \theta^\alpha )$ 
are emerged as the Faddeev-Popov (FP) ghosts stemming from the
gauge-fixing of a topological symmetry and its reducible symmetry.
Moreover, it turns out that the BRST cohomology describes the physical states 
of the superstring.

In order to understand the formalism at hand more deeply, it is useful
to recall the world-line formalism of the Chern-Simons theory in three
dimensions \cite{Witten2, Ber6} and the BF theory in arbitrary space-time 
dimensions \cite{Oda3}.  
It has been already shown that Chern-Simons theory can be
described using the world-line action \cite{Witten2, Ber6}
\footnote{Going from the first equality to the second one, we have neglected
a surface term for simplicity of the presentation. However, it should be remembered that
such a surface term in general plays an important role in the Witten type of topological 
field theories.} 
\begin{eqnarray}
S &=& \int d \tau ( P_\mu \dot{x}^\mu + l^\mu P_\mu )  \nonumber\\
&=& \int d \tau ( - x^\mu \dot{P}_\mu + l^\mu P_\mu ),
\label{WL action}
\end{eqnarray}
where the index $\mu$ takes the values $0, 1, 2$ and $l^\mu$ is a
Lagrange multiplier. In a similar way, we have also presented a 
world-line description of topological non-abelian BF theory in an
arbitrary space-time dimension and shown that its BRST cohomology
has a natural representation as the sum of the de Rham cohomology \cite{Oda3}.

When we compare the action (\ref{WL action}) with the superparticle 
action (\ref{SP action1}), we soon realize a similarity such that the superparticle 
action (\ref{SP action1}) is nothing but the world-line action (\ref{WL action})
where the space-time vector $P_\mu$ is replaced with the corresponding
spinor and the pure spinor condition is imposed on this spinor.
Thus, roughly speaking, the world-line formalism with the pure spinor
condition naturally leads to the superstring in the
pure spinor formalism in ten dimensions. 

From the viewpoint of $\beta-\gamma$ system \cite{Nekrasov, Grassi2}
\begin{eqnarray}
S = \int d^2 z  \beta_i \bar \partial \gamma^i,
\label{beta}
\end{eqnarray}
our classical action (\ref{ST action1}) of the superstring can be interpreted as 
that of the $\beta-\gamma$ system with a Lagrange multiplier enforcing a 
topological symmetry where $\gamma^i$ and $\beta_i$ is a pure spinor and its conjugate
momentum, respectively. It would be also interesting
to study the present formulation from this viewpoint further in future.

\begin{flushleft}
{\bf Acknowledgements}
\end{flushleft}

We would like to thank N. Berkovits and M. Tonin for valuable discussions. 
This work is supported in part by the Grant-in-Aid for Scientific Research (C) 
No. 22540287 from the Japan Ministry of Education, Culture, Sports, Science 
and Technology.


\end{document}